\documentclass[%
 reprint,
nofootinbib,
 amsmath,amssymb,
 aps,
pra,
floatfix,
]{revtex4-2}

\usepackage{graphicx}
\usepackage{dcolumn}
\usepackage{bm}
\usepackage{physics, tensor}
\usepackage[normalem]{ulem}

\usepackage{xcolor}

\usepackage{placeins}
\begin{document}

\preprint{APS/123-QED}

\title{Realistic Oscillon  Interactions}

\author{Angela Xue}
 \email{angela.xue@vuw.ac.nz}
 \author{Kyle  Chen}%
  \email{tche432@aucklanduni.ac.nz} 
 \author{Baylee Verzyde}%
  \email{jver961@aucklanduni.ac.nz}
\author{Peter Hayman}%
 \email{peter.hayman@auckland.ac.nz}
\author{Richard Easther}%
 \email{r.easther@auckland.ac.nz}
\affiliation{%
Department of Physics, the University of Auckland \\
New Zealand 1010
}%


\date{\today}

\begin{abstract}
Oscillons are long-lived nonlinear pseudo-solitonic configurations of scalar fields and many plausible inflationary scenarios predict an oscillon-dominated phase in the early universe.  Many possible aspects of this phase remain  unexplored, particularly oscillon-oscillon interactions and interactions between oscillons and their environment. However the primary long range forces between oscillons are gravitational and thus slow-acting relative to the intrinsic timescales of the oscillons themselves. Given that simulations with local gravity are computationally expensive we explore these effects by extracting individual specimens from simulations and then engineering interactions. We find that oscillons  experience friction when moving in an inhomogeneous background and, because oscillons in non-relativistic collisions bounce or merge as a function of their relative phases, the outcomes of interactions between ``wild'' oscillons depend on their specific trajectories.   
\end{abstract}

\maketitle

\section{Introduction}
\label{sec:introduction}

Long-lived,  pseudo-solitonic lumps \cite{coleman_aspects_1985} known as oscillons arise in non-linear field theories  and are of particular interest in the context of early universe cosmology \cite{Bogolyubsky:1976yu,Gleiser:1993pt,Copeland:1995fq, Broadhead:2005hn, Gleiser:2009ys,Gleiser:2011xj, Amin:2010jq,Amin:2010xe,Gleiser:2010qt, Amin:2010dc,Amin:2011hj,Salmi:2012ta,Zhou:2013tsa,Amin:2013ika,Mukaida:2016hwd,Lozanov:2017hjm,Cotner:2018vug,Lozanov:2019ylm}. In particular, current cosmological data favors inflationary models with potentials that grow sub-quadratically \cite{Planck:2018jri,BICEP:2021xfz,Tristram:2020wbi,Easther:2021rdg,ACT:2025tim}, a necessary condition for the existence of oscillonic solutions. In addition, these models typically support post-inflationary resonance \cite{Lozanov:2016hid, Lozanov:2017hjm}, which is required for the formation of oscillons. Consequently, it is plausible that the early universe passes through an oscillon dominated phase, such as that described for axion monodromy models \cite{Amin:2011hj}. 

Oscillons are supported by nonlinear interactions in the field sector and cosmological oscillons are typically studied in simulations in which the overall space expands without local gravitational interactions \cite{Felder:2000hq,Felder:2007kat,Sainio:2009hm,Easther:2010qz,Figueroa:2021yhd}. In these scenarios there are no significant inter-oscillon forces and thus few interactions between them.  However, gravitational clustering will cause the growth of structures in the early universe \cite{Jedamzik:2010dq,Easther:2010mr,Lozanov:2019ylm,Eggemeier:2020zeg,Eggemeier:2021smj} greatly increasing the likelihood of oscillon-oscillon interactions if they live for several Hubble times. 

Interactions between oscillons are largely unexplored. Most work to date takes idealized  profiles as initial conditions rather than ``naturally-occurring'' or ``wild'' oscillons \cite{Hindmarsh:2006ur,Hindmarsh:2007jb,Amin:2020vja}, or is performed within a Schr\"{o}dinger-Poisson-style approximation which captures gravitational interactions but is non-relativistic \cite{Amin:2019ums}. Here, we take a complementary approach, extracting oscillons formed in cosmological simulations, boosting them towards each other to force interactions and performing well resolved simulations of the subsequent collisions. This allows us to look at  interactions between cosmologically realistic oscillons while sidestepping large-scale simulations with both local gravity and the full field dynamics.

In more detail, our strategy is as follows. We take an inflaton potential (the generalized monodromy potential) that is known to support oscillons for a range of cosmologically relevant parameter choices and simulate the post-inflationary evolution of the inflaton. We used the well-established Einstein-Klein-Gordon solver \texttt{ClusterEasy} \cite{felder_clustereasy_2008}, self-consistently solving for both the inflaton field and the scale factor in a Friedmann-Robertson-Walker universe.   We initialize the field with the standard Bunch-Davis power-spectrum and evolve until the universe is well populated with oscillons.  We then  algorithmically identify and label each oscillon, selecting a few clean samples. These are used to run new simulations using pairs of extracted oscillons boosted towards each other to generate collisions and a single oscillon interacting with planar waves to study drag.

Two subtleties arise. First,  an oscillon occupies a small fraction of the lattice volume and we interpolate the configuration to higher resolutions as a proxy for a full adaptive mesh refinement scheme, e.g. Ref.~\cite{zhang_amrex_2019}. Second, in the absence of local gravity the generated oscillons are approximately stationary.  Oscillons are typically much larger than their Schwarzschild radii so gravitationally induced collisions are likely to be sub-relativistic. This is in contrast to previously studied ultra-relativistic collisions \cite{Amin:2014fua}, and  2D oscillons (which are simpler numerically but behave similarly to their 3D analogues) interacting at velocities which are a significant fraction of the speed of light \cite{Hindmarsh:2006ur,Hindmarsh:2007jb}.   

Given that we are interested in low-speed collisions it might seem that we could na\"\i vely apply Galilean shifts to the extracted field data. However, we will see that the boosts are nontrivial since oscillons are emergent structures and there is nonlinear relationship between the nominal boost velocity, the resulting speed of the oscillon and the field data.  

The overall goal of this paper is to describe and verify this approach to studying oscillon dynamics. 
We examine drag induced by interaction with background field perturbations (which to our knowledge has not been studied even with idealized profiles) by embedding a single oscillon in a background with planar waves.  We then look at a range of representative interactions, reproducing known results (e.g., mergers and bounces depending on relative phase \cite{Amin:2019ums,Amin:2020vja}) for wild oscillons. We show that, as a consequence, the outcome of collisions of mismatched oscillons will depend of their  starting positions and  speeds. 
 
The paper is organized as follows: Section \ref{sec:simulations} outlines the underlying model and the generation of oscillons. Their identification and extraction is described in Section \ref{sec:extractions}, along with how the details as to how collisions are arranged and simulated. Section \ref{sec:drag} discusses the coupling between small perturbations and a moving oscillon Section \ref{sec:collisions} and we conclude and remark on future directions in Section \ref{sec:conclusions}. 

\section{The Model}
\label{sec:simulations}

We work with the generalized monodromy model \cite{Amin:2011hj}, in which the inflaton field $\phi$ has the potential 
\begin{eqnarray}
\label{eq:sec1:potential}
	V(\phi) = \frac{m^2M^2}{2\alpha}\bigg[\bigg(1+\frac{\phi^2}{M^2}\bigg)^{\alpha}-1\bigg]\, .
\end{eqnarray}
This model is  motivated by axion monodromy constructions in string theory \cite{McAllister:2008hb,Silverstein:2008sg} and supports both oscillon solutions and the parametric resonance needed to  spawn them \cite{Amin:2011hj}. Its precise predictions are at odds with the Planck + BICEP microwave background dataset  \cite{BICEP:2021xfz}. However, they are a better match to the more recent ACT results \cite{ACT:2025tim}, especially if there is a long matter dominated phase after inflation, reducing the number of e-folds $N$ before the end of inflation at which cosmological modes leave the horizon. In any case the scenario is well-studied and none of the techniques we develop here  depend on the specific choice of potential.

Oscillonic solutions require that the potential scales as a sub-quadratic power-law away from its minimum \cite{amin_flat-top_2010}, which is true  when  $\alpha<1$. The parameter $M$ sets the width of the quadratic minimum and it is  convenient to use the dimensionless $\beta\equiv M_{\text{Pl}}/M$, where the (reduced) Planck mass is $M_{\text{Pl}}=1/\sqrt{8\pi G}$.  We initialize our cosmological simulations at the end of inflation and require that the model reproduces the observed density perturbations (following Ref.~\cite{Amin:2011hj}) which fixes  $m^2$  for a given $\alpha$ and $\beta$. 

\begin{figure}[t]
    \centering
    \includegraphics{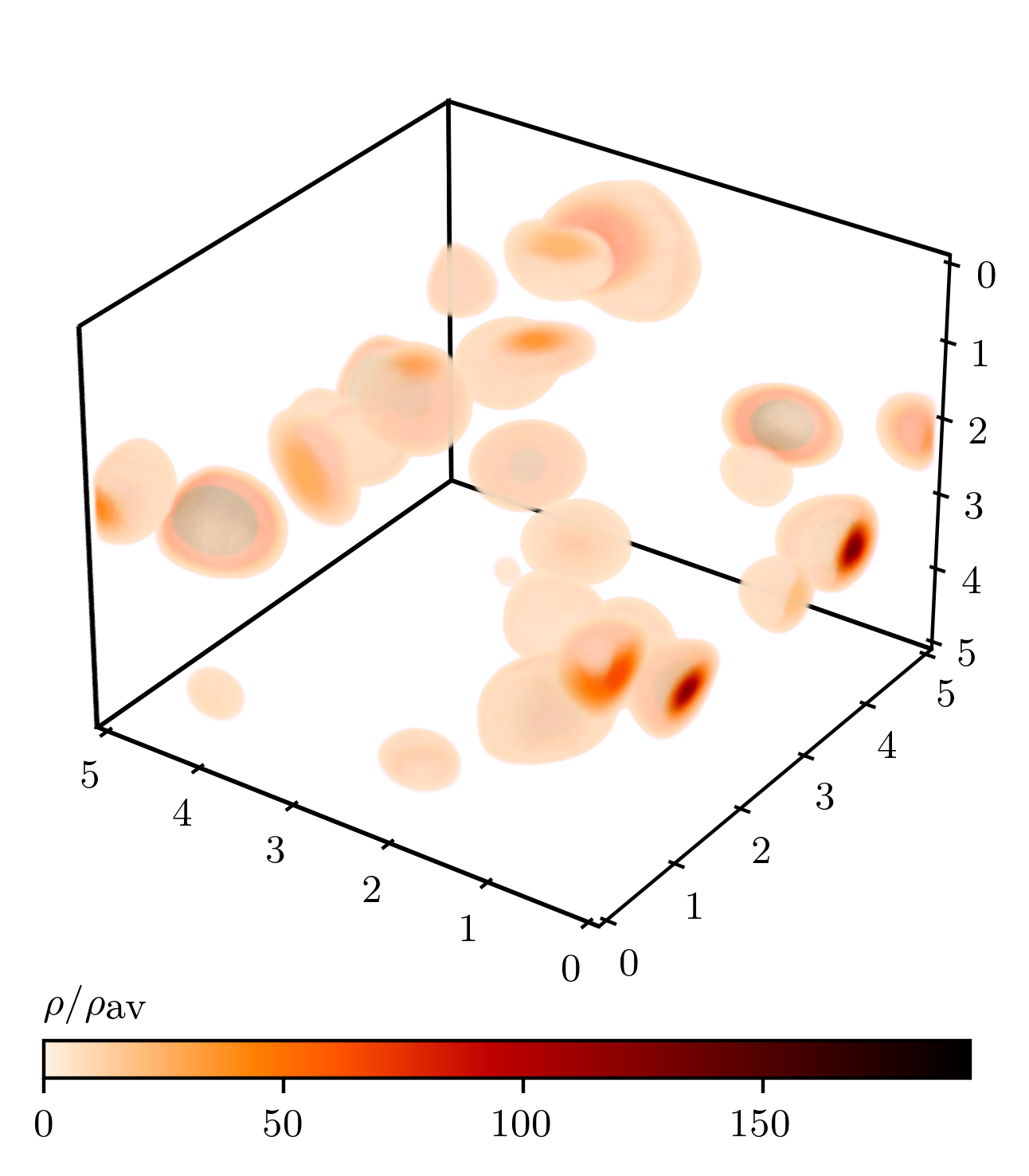}
    \caption{The relative energy density, $\rho/\rho_\text{av}$, is plotted for $(\alpha,\beta)=(0.5,50)$ at $2.37$ e-folds into the simulation, showing a healthy population of oscillons. }
    \label{fig:universe}
\end{figure}

Our equations of motion are the coupled Klein-Gordon and second Friedmann equations 
\begin{align}
   \ddot\phi + 3H\dot \phi -\frac 1 {a^2} \nabla^2 \phi + \frac{\partial V(\phi)}{\partial \phi} &= 0, \label{eq:sec1:EOM1}\qquad  \\
   \frac{\ddot{a}}{a} + \frac{4\pi G}{3}(\rho + 3 p) &= 0, \label{eq:sec1:EOM2}
\end{align}
respectively, and the system obeys the constraint expressed by the first Friedmann equation
\begin{equation}
\label{eq:sec1:EOMconst}
    H^2 = \frac{8\pi G}{3}\rho \, .
\end{equation}
As usual, $a$ is the scale factor, $H = \dot a / a$ is the Hubble constant. Overdots denote differentiation with respect to time $t$, and $\rho$ and $p$ are respectively the energy density and pressure in the scalar field.  We perform our simulations with the widely-used \texttt{ClusterEasy} \cite{felder_clustereasy_2008}.  The simulations begin at the moment where the field velocity first becomes zero.\footnote{Strictly speaking, this condition is applied to the rescaled field variable used internally by \texttt{ClusterEasy}  not the physical field, but the difference is immaterial.}  A representative oscillon population a little less than 3 e-folds from the beginning of the simulation is shown in Figure \ref{fig:universe}. Length scales in plots are given in units of $1/(am)$.

\texttt{ClusterEasy} uses a re-scaled set of variables and coordinates, 
\begin{equation}
    \label{eq:progUnits}
    \phi_\text{pr} := A a^r \phi, \quad x^i_\text{pr} := B x^i, \quad \frac{\partial}{\partial t_\text{pr}} := \frac 1 {B a^s}\frac{\partial}{\partial t} \, .
    \end{equation}
In our case these are
\begin{align}
   A = \frac 1 {\phi_0}, &\quad B = mM\phi_0^{-1 + \alpha}, \notag \\
   r = \frac 3 {1 + \alpha}, &\quad \text{and} \quad s = 3\frac{1 - \alpha}{1 + \alpha}    
\end{align}
given that the dominant term of our potential  is
\begin{equation}
\label{eq:sec1:leadingPot}
    V\sim\frac{m^2M^2}{2\alpha}\phi^{2\alpha}.
\end{equation}

The dependence of oscillon production on $\alpha$ and $\beta$ is known \cite{Amin:2011hj}. We perform initial simulations with $256^3$ grids. The initial inhomogeneities are given by the standard post-inflationary Bunch-Davis  spectrum, with a momentum cutoff at   $k/B=16$ for numerical stability \cite{polarski_semiclassicality_1996}.

\section{Extracting and Boosting Oscillons}
\label{sec:extractions}

\begin{figure}[t]
\centering
    \includegraphics[width=0.48\textwidth]{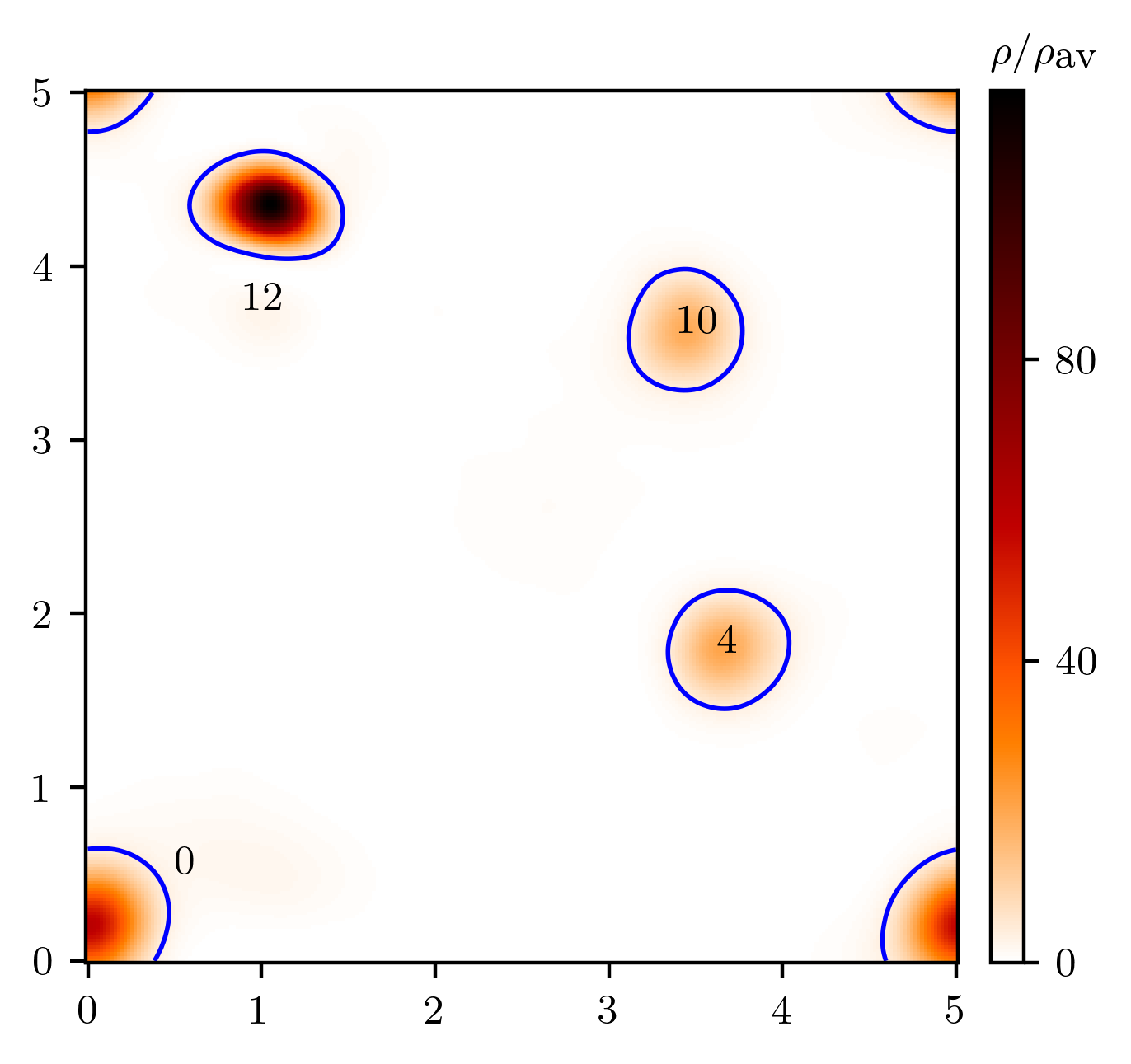}
    \caption{A slice showing the relative energy density $\rho/\rho_\text{av}$ from the plot in Figure \ref{fig:universe} with labeled oscillons. The 0th oscillon spans all corners of the grid due to  the periodic boundary condition. The blue contour marks the $\rho/\rho_{\textrm{av}}=4$ threshold.}
    \label{fig:labels}
\end{figure}

Once the oscillons in a simulation are well-established  we  extract them to build a ``library'' of realistic oscillon configurations for a given $\alpha$ and $\beta$. The oscillons are typically well separated at formation and the density contrast grows as the simulation continues. The physical size and peak density  of an oscillon is roughly fixed so their comoving size decreases and the contrast between the oscillon and its surrounding increases as the universe expands. There is thus a tradeoff between  well-separated oscillon and spatial resolution, given that we are working with a fixed grid. We allowed roughly $\sim0.2$ Hubble times to elapse after oscillon formation before extraction. 

We identify and label oscillons by assigning unique integers to each gridpoint where $\rho/\rho_\text{av} \ge 4$ and then iteratively replace each number by its lowest valued neighbor, taking into account the periodic boundary conditions. This gives a ``mask'' that defines the oscillon locations and uniquely labeling each one. A slice though a labeled grid is shown in Figure~\ref{fig:labels}.  We also performed  manual checks to exclude rare situations with interactions between oscillons or localized transients.

\begin{figure}[tbp]
\centering
\includegraphics{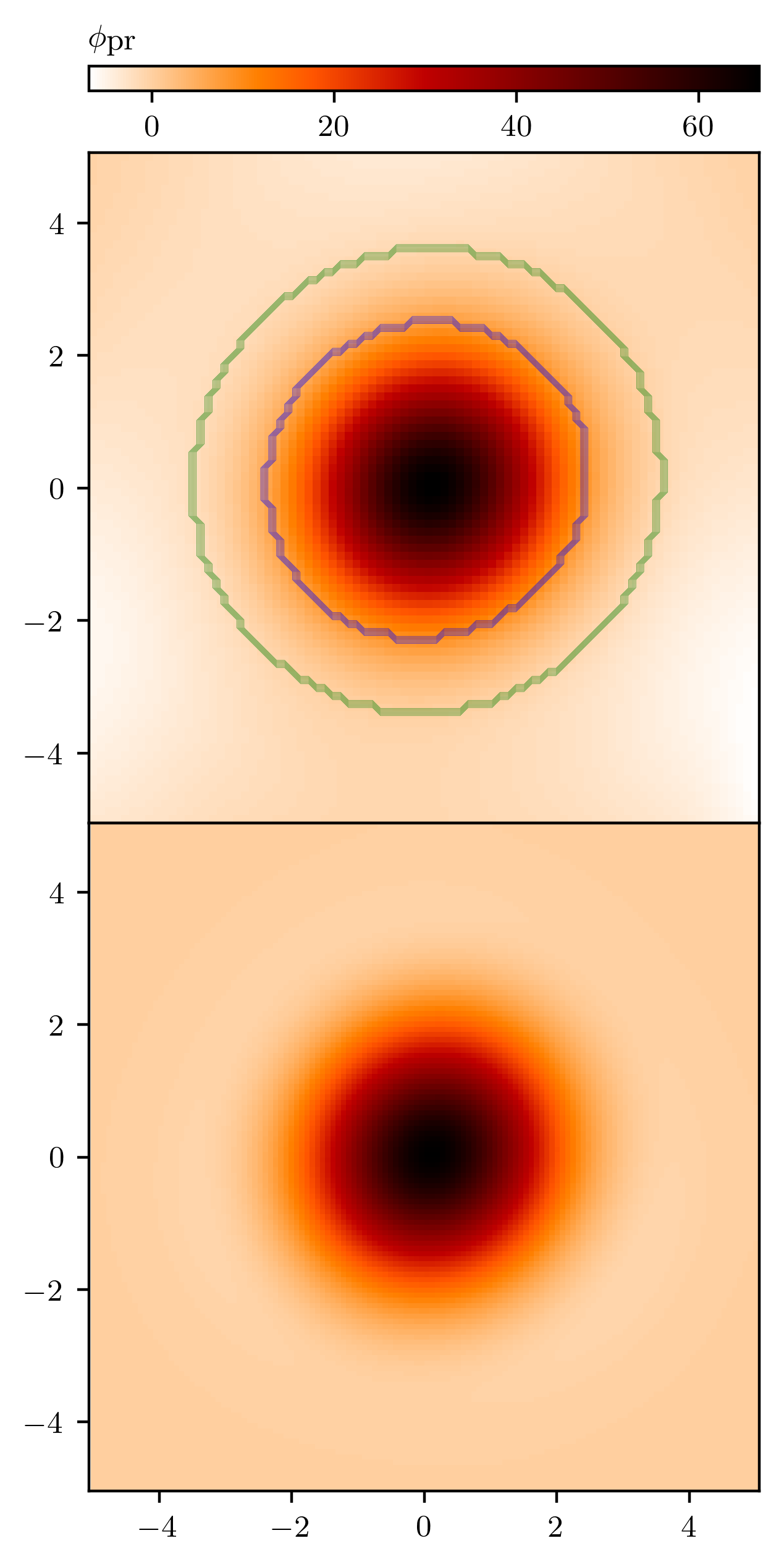}
    \caption{(Top) A slice of through an oscillon from a  simulation with $\alpha=0.05$, $\beta=25$ at 1.69 e-folds. The blue contour outlines the initial mask; the green contour shows the full extracted region. (Bottom) The same configuration on a  $128^3$ grid with the fluctuations blurred.} 
    \label{fig:refine}
\end{figure}

We extract  $\phi$ and $\dot \phi$ at the location of each oscillon with an extra region of ``padding'' to which we apply a Gaussian blur to remove  background fluctuations. For our setup the extracted regions were typically $\sim\mathcal{O}(80^3)$ grid points across and we used linear interpolation to increase the resolution if needed. An example is shown in Figure \ref{fig:refine}.  Given the modifications to their outer regions we do not expect that the subsequent evolution of the extracted oscillons will exactly match that in their original habitat. However, the qualitative behavior  of an extracted oscillon matches that seen in the primary simulation.  

Ultra-relativistic collisions have been studied by Amin {\em et al.\/} \cite{Amin:2014fua}. However,  collisions driven by gravitational interactions will be  sub-relativistic  given that typical oscillons are much larger than their Schwarzschild radii. 
Oscillons are emergent structures of the inflaton field rather than point-like particles, so we look  at their energy densities and linear momenta to deduce their  velocities.

Consider the overall linear momentum density $P_i$ of the inflaton field defined through its stress-energy density $T_{\mu\nu}$ with a mostly-plus metric signature,
\begin{equation}
    P_i\equiv\tensor{T}{^{0i}}=-\frac{1}{a^2}\dot\phi\partial_i\phi.
    \label{eq:defP}
\end{equation}
For an oscillon moving at non-relativistic speeds its energy density is a good approximation to its mass. We define its velocity as the ratio of its momentum to its (approximate) mass, 
\begin{equation}
   v_i := \frac{(P_i)_\text{av}}{\rho_\text{av}} \, ,
\end{equation}
where $(P_i)_\text{av}$ and $\rho_\text{av}$ are  the average momentum and energy density calculated across the oscillon volume. 

We need to boost the extracted oscillons in order to induce interactions. Given our interest in non-relativistic collisions and because the time-slicing of  a lattice simulation breaks Lorentz invariance we consider  Galilean boosts
\begin{equation}
\label{eq:boost}
\phi(t, \vec{x}) \to \phi(t, \vec{x} - \vec{u}(t-t_0))\, .
\end{equation}
Setting $t_0=0$ this gives a shift in the time derivative  
\begin{equation}
\dv{\phi(t, \vec{x} - \vec{u}t)}{t}= \pdv{\phi(t, \vec{x} - \vec{u}t)}{t} - \vec u\cdot\nabla\phi(t, \vec{x} - \vec{u}t),
\label{eq:boostdt}
\end{equation}
which moves  the initial data by
\begin{equation}
\dot\phi_0 \to \dot\phi_0 - \vec u\cdot\nabla\phi_0.
\label{eq:compBoost}
\end{equation}
Applying the boost \eqref{eq:compBoost} leads to a shift in momentum density
\begin{equation}
    P_i \to P_i - \frac 1 {a^2}(\vec{u}\cdot\nabla\phi_0)\partial_i\phi_0.
    \label{eq:boostP}
\end{equation}

\begin{figure}[t]
\centering    \includegraphics[width=0.48\textwidth]{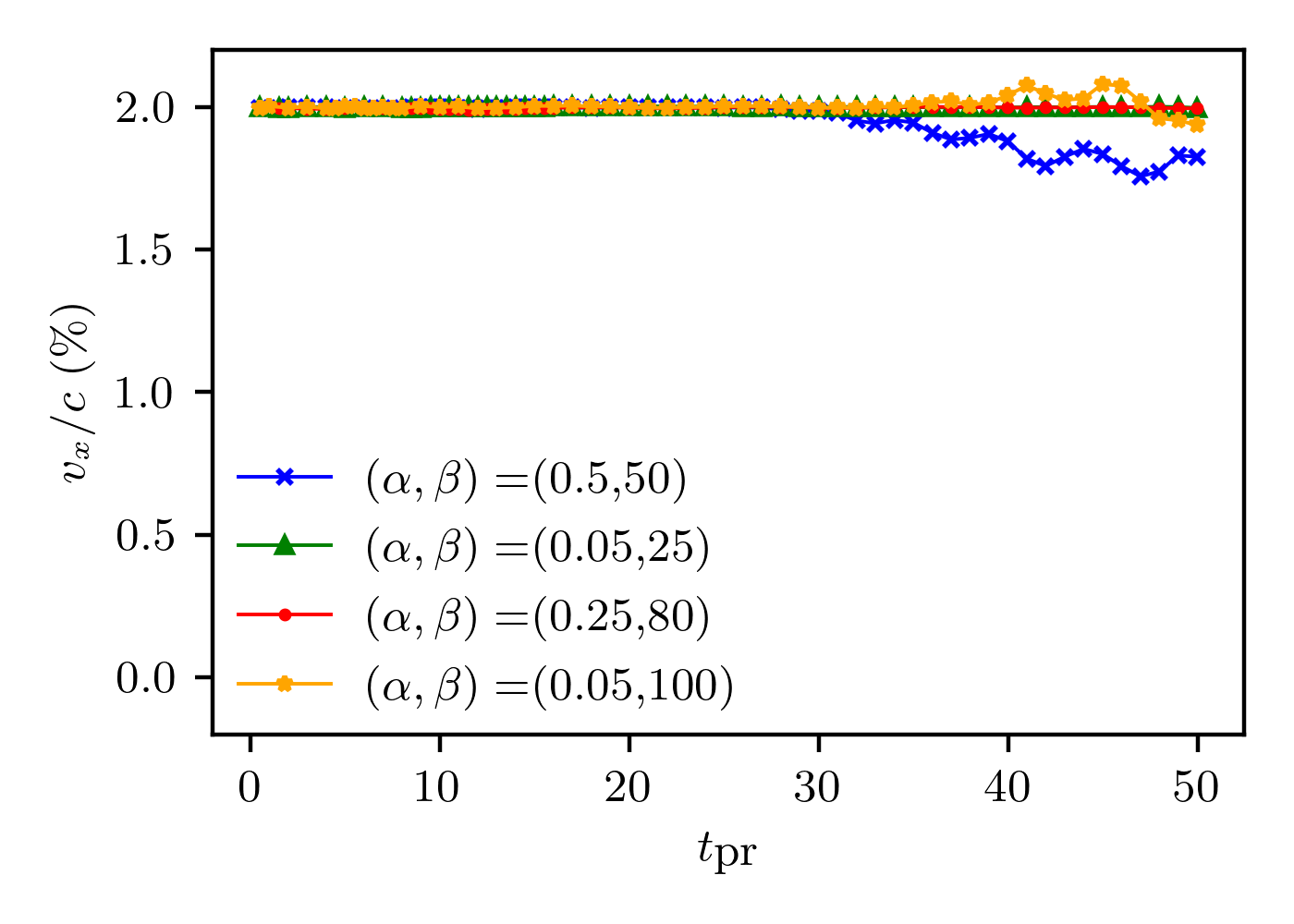}
    \caption{The center of energy speeds of oscillons that were given desired boosts of $v_x\sim0.02c$ and $v_y\sim v_x\sim0$. Each oscillon is extracted from different simulated universes and made to evolve in an empty and non-expanding universe.  }
    \label{fig:compare speeds}
\end{figure}

Broadly speaking, oscillons are ``jogging on the spot'', with a fixed (or slowly varying) spatial envelope $f(r)$ and rapid temporal oscillations,  
\begin{equation}
\label{eq:template}
\tilde\phi \sim f(r) \cos{\omega t} \, .
\end{equation}
Consequently, the change in speed induced by a Galilean boost with the form of equation~\ref{eq:boost} will depend on its phase. Moreover, wild oscillons are not spherically symmetric so a boost in one direction can change the velocity in all three directions. Figure~\ref{fig:compare speeds} shows the measured speeds of four boosted oscillons, each drawn from simulations with different values of $\alpha$ and $\beta$.  We set a velocity $\vec{v}$ for an oscillon and then iteratively solve for the $\vec{u}$ that delivers it. We find that $\vec{u}/c$ is much larger than the resulting $\vec{v}/c$ and has three non-trivial components. For example with $(\alpha,\beta)= (0.5,50)$ an oscillon velocity $\vec{v}/c \sim 0.02$ required $\vec{u}/c = (0.4138, -0.0769, 0.3268)$ while for $(\alpha,\beta)= (0.05,100)$,  $\vec{u}/c =(0.329, 0.1762, -0.1198)$.

The clearly relativistic shift parameters indicate that even at non-relativistic speeds (such as those caused by local gravitational attraction) the internal dynamics of oscillons are  relativistic. In addition, the outcome of the Galilean boost depends on the phase of the oscillon and its asphericity, which tends to decrease as it evolves so the value of $\vec{u}/c$ needed for a given boost $v$ will vary significantly with time. Nevertheless, for our present work these Galilean boosts are  empirically effective  so we use them in what follows.


\begin{figure*}[!tb]
    \centering
    \includegraphics[width=\linewidth]{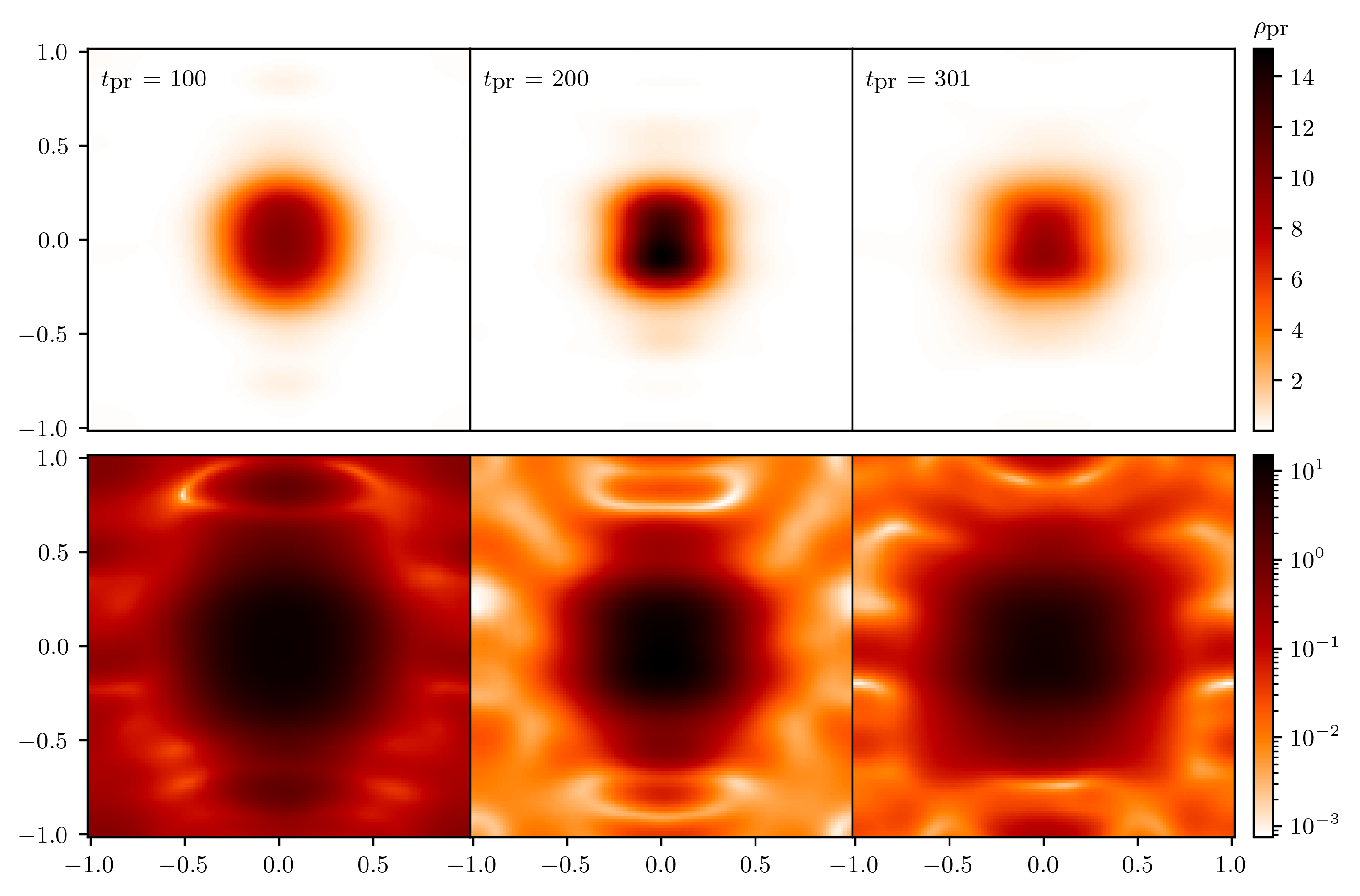}
    \caption{An isolated oscillon subject to an incoming planar wave $(\alpha,\beta)=(0.5,50)$. Top is a linear scale to see the oscillon in detail, while bottom is the same plot in a logarithmic scale to bring out the background radiation.}
    \label{fig:dragFields}
\end{figure*}

\begin{figure}
    \centering
    \includegraphics[width=0.48\textwidth]{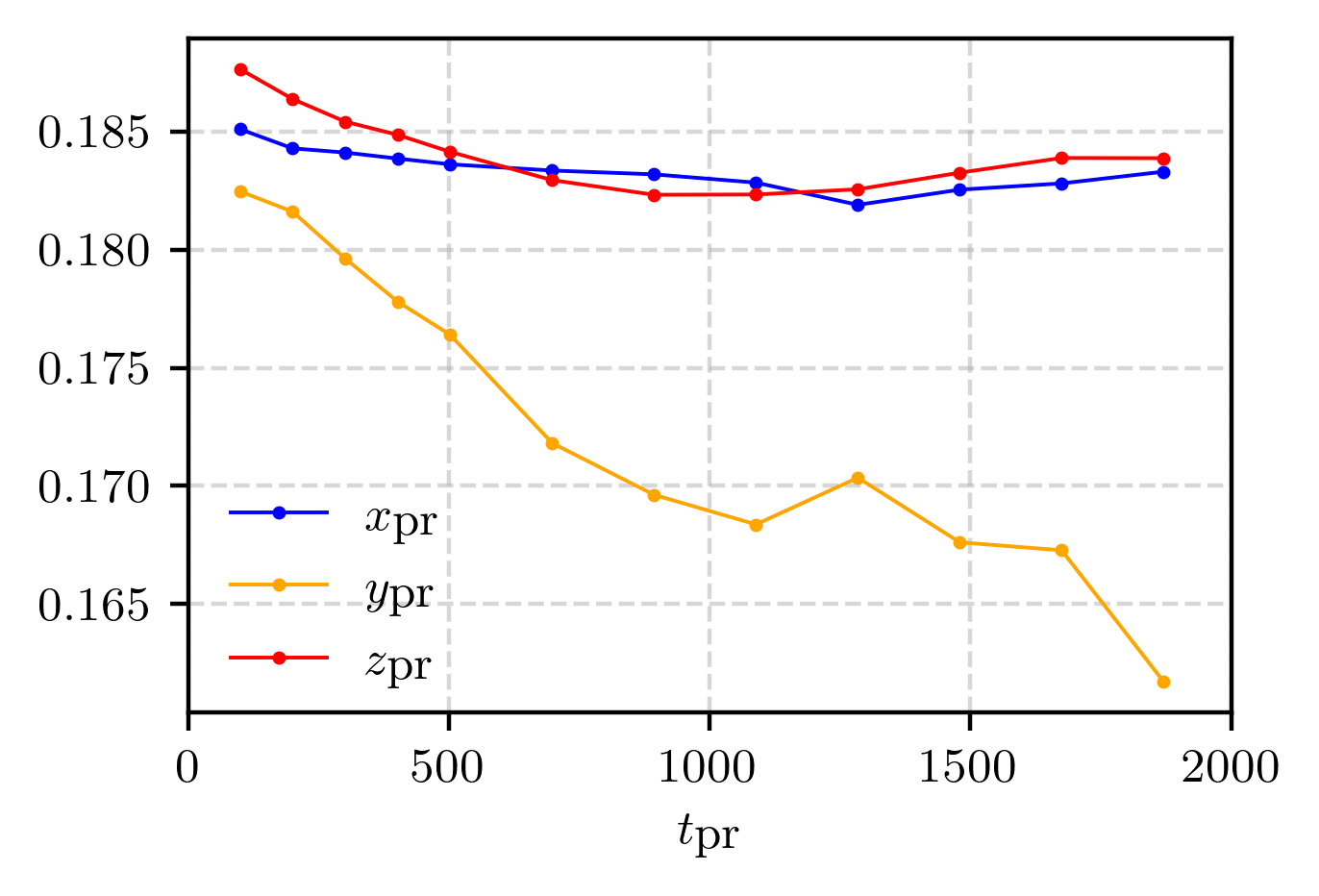}
    \caption{The $x,y,z$ position of the oscillon from Figure \ref{fig:dragFields} as a function of time. }
    \label{fig:dragPositions}
\end{figure}

\section{Oscillons and Background Waves}
\label{sec:drag}

A post-resonance oscillon-dominated universe contains a mix of oscillons and propagating perturbations. Consequently, we begin by looking at individual oscillons in backgrounds containing propagating waves.  The time-averaged momentum flux on a stationary oscillon is expected to be zero but a moving oscillon will face an effective ``headwind''. However, given that we are focusing on local interactions we consider a scenario in which the oscillon is initially at rest in a background of propagating scalar waves. 

With Hubble expansion turned off (and $a=1$ for convenience) the Klein-Gordon equation reduces to 
\begin{equation}
    \ddot{\phi} - \nabla^2 \phi + m^2 \phi =0 \, 
\end{equation}
for small perturbations. This has plane-wave solutions 
\begin{equation}
    \phi = A \exp{i(-\omega t + \vec{k}\cdot \vec{x})} \, 
\end{equation}
when $k^2 + m^2 - \omega^2 =0$.

Figure \ref{fig:dragFields} shows an initially stationary isolated oscillon in a planar wave background; the amplitude of the waves is far less than the height of the oscillon. Figure \ref{fig:dragPositions} plots the oscillon position;  it is  stationary in $x$ and $z$ but is pushed at a roughly constant velocity in the $-y$ direction, the propagation direction of the waves.    The amplitude of the waves is a few percent of the maximum oscillon amplitude.

\begin{figure*}[tb]
\centering
    \includegraphics{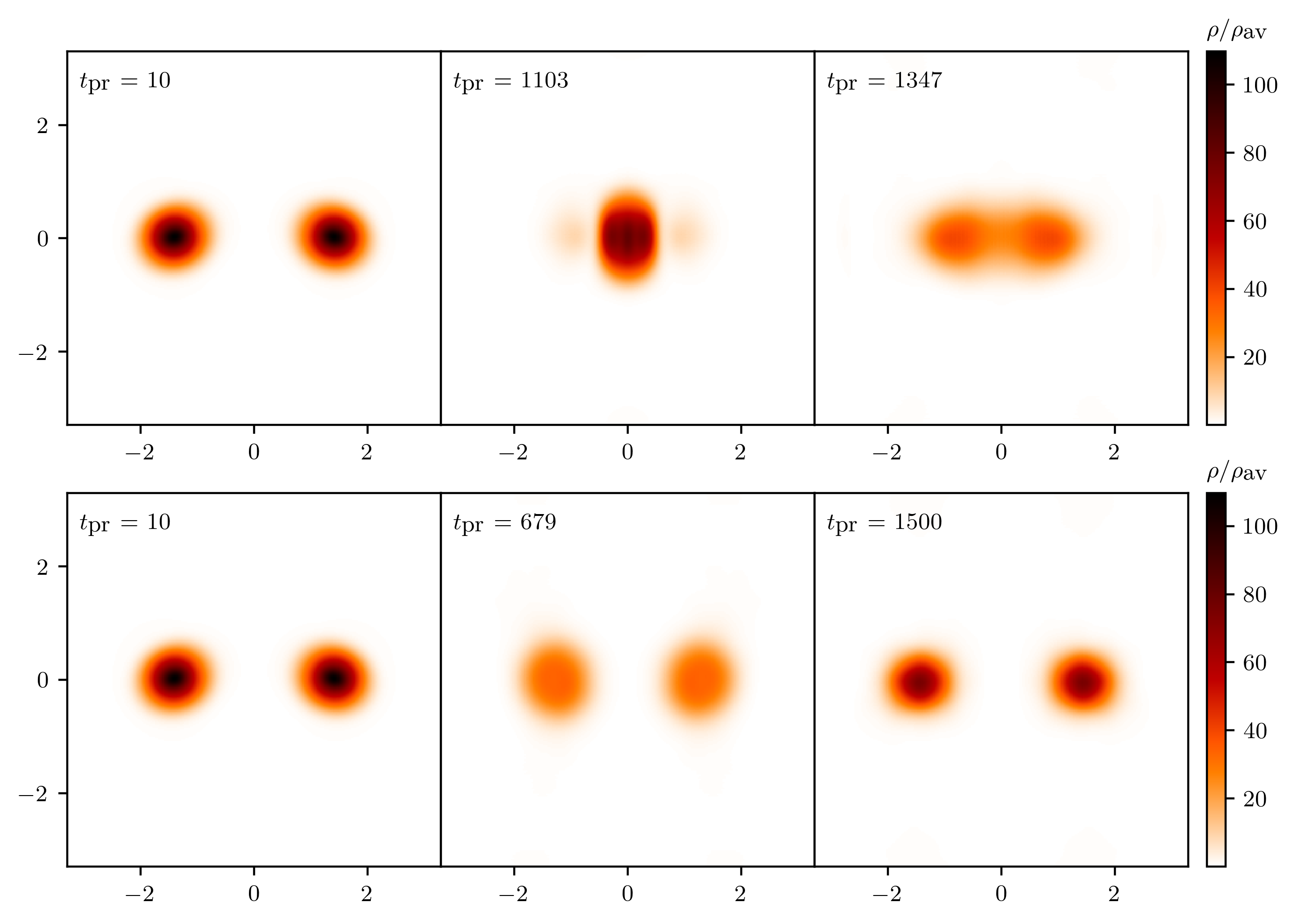}
    \caption{(Top) A merger between an oscillon and its mirrored copy, each traveling at $0.02c$, for a relative speed of $0.04c$. The left-most image shows the oscillons are well before the merger, the middle and right images show the  most contracted state and maximal rebound, respectively. (Bottom) As above, but out-of-phase; the closest approach in the middle image, and the right image shows them after rebounding. Both plots made with the same oscillon from a $(\alpha,\beta)=(0.05,25)$ simulation.}
    \label{fig:mergerandbouncer}
\end{figure*}

An oscillon moving in an otherwise empty universe does not significantly decelerate, as seen in Figure~\ref{fig:compare speeds}. However, if the universe contains a background of scalar field waves -- as it will if the oscillon population has been produced via resonance -- then  oscillons moving in gravitational potentials will experience drag.  The  consequences of this for the oscillon dominated phase are unclear -- however, oscillons approaching each other via their mutual gravitational attraction will not ``free fall''.

\section{Oscillon Collisions}
\label{sec:collisions}

Our approach allows us to stage interactions between any pair of oscillons. To induce interactions we create an initial configuration consisting of two well-separated  oscillons, moving towards each other with the same speed.  We  evolve this  configuration with  \texttt{ClusterEasy} in a fixed background, using either an $128^3$ or a  $256^3$ grid.  

We begin by confirming the phase-dependence of the outcome of  a collision by  colliding oscillons with identical copy of themselves, for in-phase and maximally out-of-phase starting conditions. The latter is achieved by multiplying its field and derivative  by $-1$. Representative results are shown in Figure~\ref{fig:mergerandbouncer}. All in-phase collisions give a merger and out-of-phase collisions give a bounce, as expected \cite{Amin:2019ums,Amin:2020vja}.

\begin{figure}[tb]
    \centering    \includegraphics[width=0.89\linewidth]{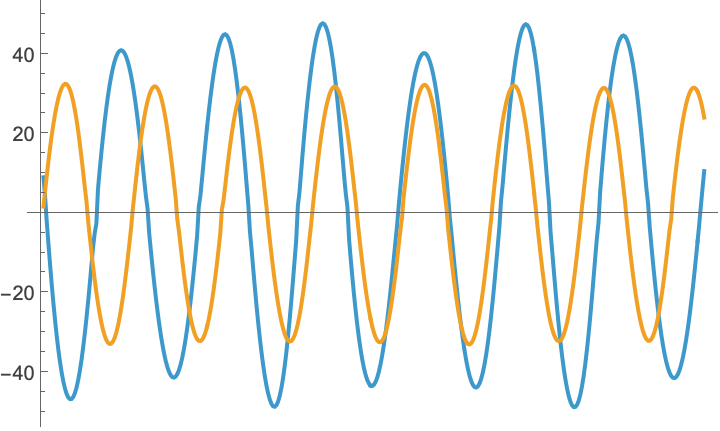}
    \caption{The maximal field amplitudes for the of mismatched oscillons shown in Figure~\ref{fig:merge} shortly before they interact. The different frequencies are clearly visible, and the larger oscillon has a second, long period oscillation in its amplitude.}
    \label{fig:modulation}
\end{figure}

\begin{figure}[p]
    \centering    
\vspace{3.5mm}    
    \includegraphics[width=0.88\linewidth]{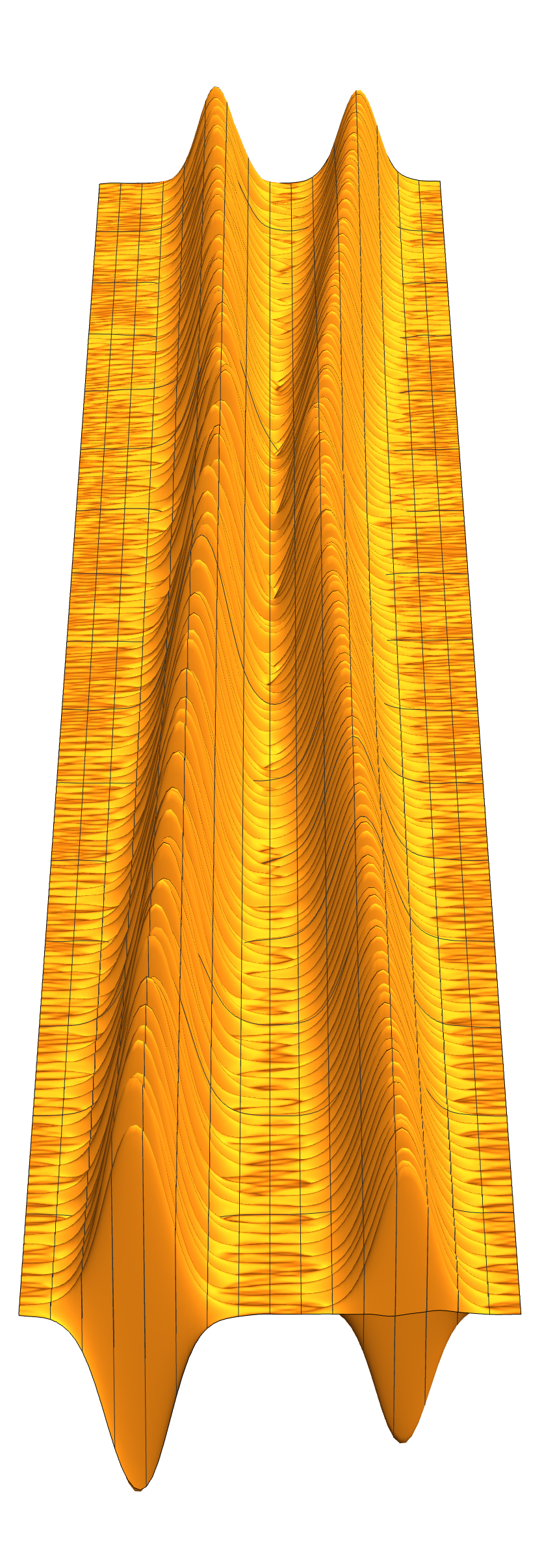}
    \caption{Bounce between two mismatched oscillons, each with a speed of $0.01c$. Time runs vertically.}
    \label{fig:bouncer}
\end{figure}

\begin{figure}[p]
    \centering    \includegraphics[width=0.88\linewidth]{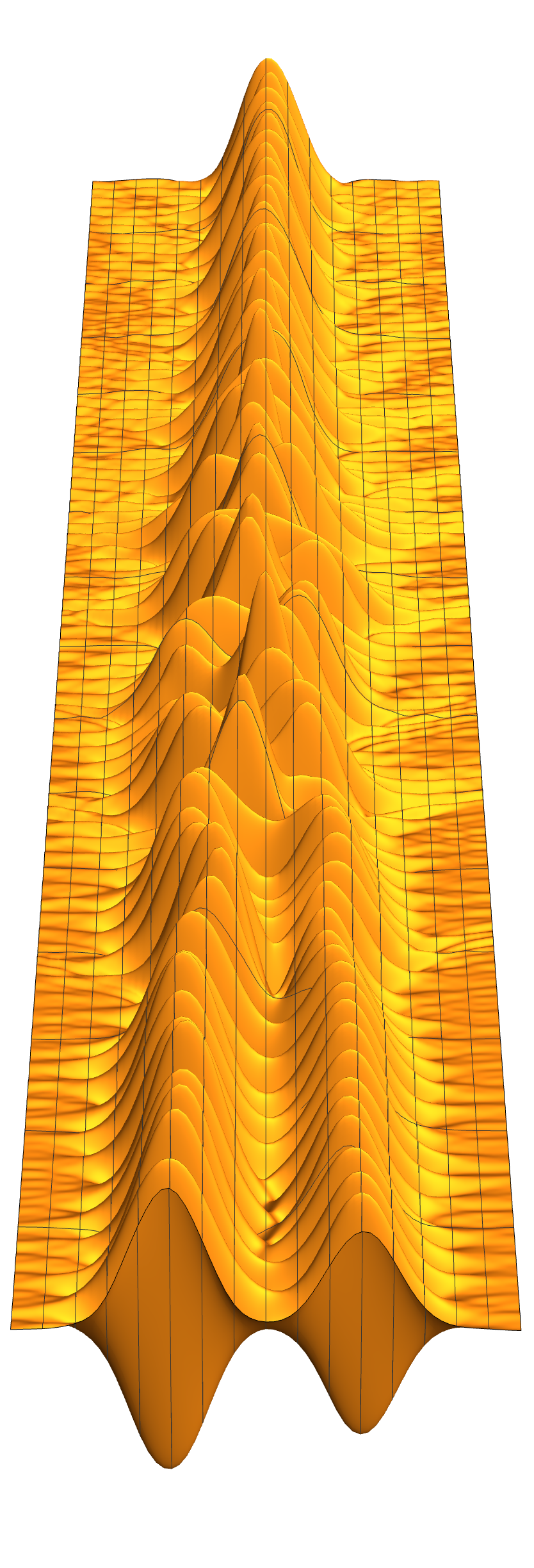}
    \caption{Detail of merger between two mismatched oscillons, each with a speed of $0.015c$. Time runs vertically.}
    \label{fig:merge}
\end{figure}

By construction, ``matched pairs'' of oscillons have a well-defined mutual phase since the oscillon is essentially interacting with its  clone. However, the frequency of each member ($\omega$ in   equation~\ref{eq:template})  of an ensemble of solitons is unique and the overall envelope $f(r)$ can also undergo a slower modulation, as shown in Figure~\ref{fig:modulation}. As a result, the outcome of an interaction between two different oscillons can be expected to depend on their closing speed and initial separation as these determine their relative phase when they actually interact. 

Representative interactions between mismatched oscillons are shown in Figures \ref{fig:bouncer} and \ref{fig:merge}. In both cases the left hand oscillon is the more massive member of the pair. The  closing velocities prior to the bounce are $0.01c$ and $0.015c$ for the merger. In the case of the bounce the two oscillons barely touch before their mutual recoil and  it occurs in essentially a single oscillation. The collision is ``particle like'' -- the smaller member of the pair has a greater recoil speed  than its larger counterpart.

\begin{figure}[tb]
    \centering    \includegraphics[width=0.89\linewidth]{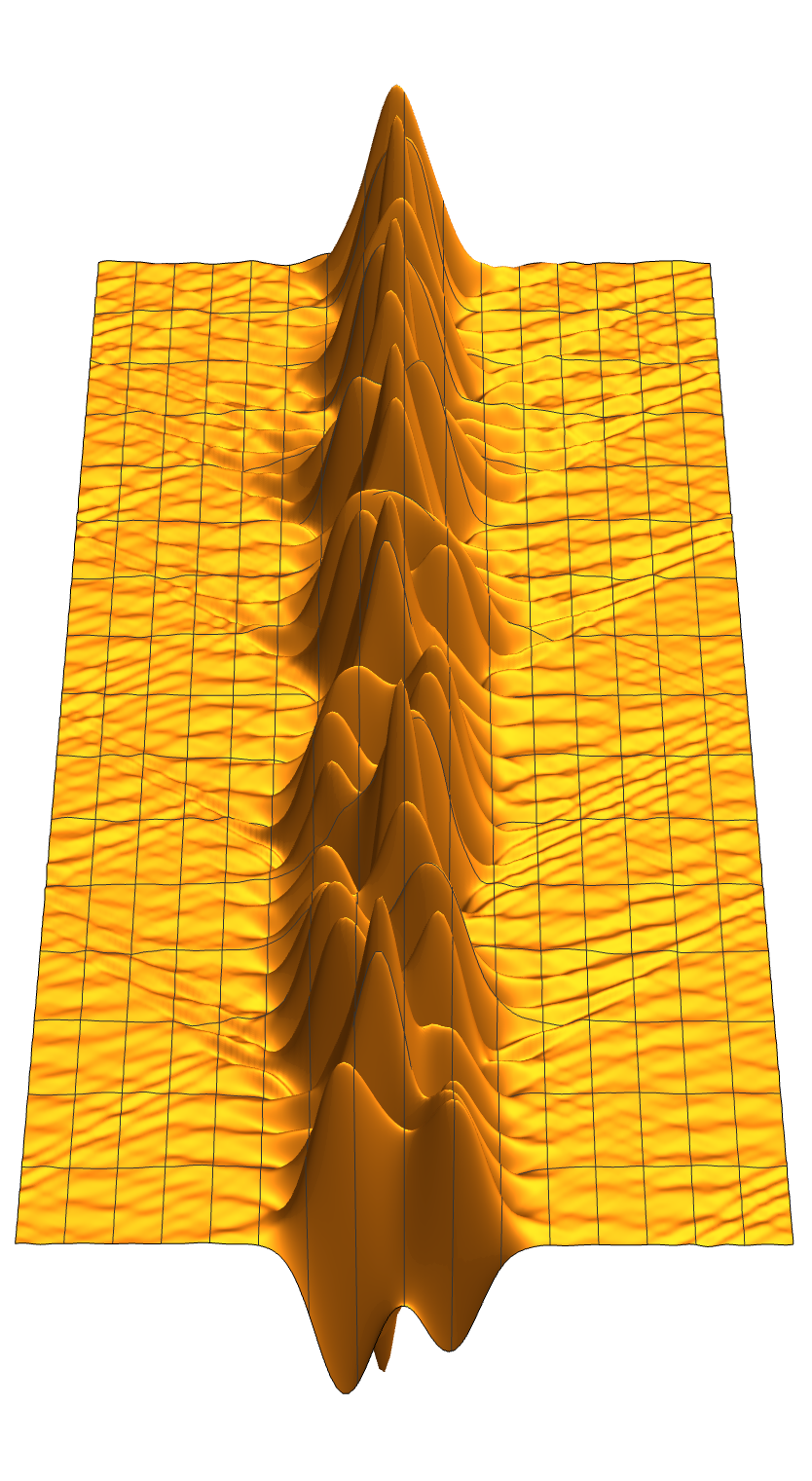}
    \caption{Detail of merger in Figure~\ref{fig:merge}. Scalar waves propagating away from the interaction are  visible.}
    \label{fig:wavy}
\end{figure}

Conversely, the merger is inelastic  and the interaction  is followed by a ringdown as the combined oscillon undergoes large excursions as it settles into a single, more symmetric profile. In this case we  see scalar waves propagating away from the interaction region so the mass-energy of the merged oscillon will be less than that of its  precursors. Note  that waves reenter from the opposite side of the box due to the periodic boundary conditions. 

We built a set of sample interactions by colliding mismatched pairs of oscillons with speeds of $0.01$, $0.015$ and $0.02c$, each from five different starting positions. In addition we  began each configurations with one of the oscillons flipped -- giving 60 collisions in total.\footnote{The specific scenario from which the oscillons were drawn had $(\alpha, \beta) = (0.3,70)$.}   Interestingly, there is no clear pattern in the outcomes -- in some cases both the orginal and flipped collisions give the same results and some pairs of oscillons are apparently more likely to merge than others. It is  clear that interactions between realistic wild oscillons have a rich and currently unexplored range of properties.


\section{Conclusion and Discussion}
\label{sec:conclusions}

We  carried out the first simulations of low-speed collisions between realistic oscillons. Unique, individual oscillons produced in a post-inflationary  simulation were extracted and non-relativistically boosted towards one other. When identical oscillons interact their relative phases determines whether they merge or bounce.  Conversely, the outcome for mismatched oscillons is essentially stochastic since their internal frequencies are different and they move in and out of phase.  

We performed na\"\i ve Galilean boosts. However the effective velocity is a convolution of the Galilean shift with the  oscillon profile and  the resulting velocity depends on the phase of the oscillon and any anisotropy in the initial configuration, as well as the nominal boost parameter. Moreover, oscillons  relax into more spherical configurations  as they evolve so the outcome  also depends on how long the oscillation  evolves before the boost is applied. 

Small shifts in the oscillon velocity  required a large boost parameter. This apparent paradox is resolved by recalling that while the oscillons are massive and non-relativistic  their internal dynamics involve high energy modes. The underlying field theory (in a static background) is fully Lorentz invariant and implementing a Lorentz boost that  mixes data from different time-steps in the simulation is  an obvious extension of this work.   

Despite their relativistic interiors the oscillons  largely behave  like classical particles. An oscillon  accelerates in a background of anisotropic waves and, we infer, would decelerate if it was moving in a statistically isotropic background of random waves. Likewise when a smaller oscillon recoils off a larger one the interaction is reminiscent of a low-speed and largely elastic collision. 

Interactions between in-phase of maximally out-of-phase oscillons proceed as expected, with mergers and bounces respectively. However,  interactions between pairs of mismatched wild oscillons (i.e. oscillons drawn from a simulation of the post-resonance universe) are effectively stochastic given that each oscillon has a unique internal frequency. Consequently, their relative phase at the moment of interaction cannot easily be set in advance.

This work opens  multiple avenues for future investigations.  We  have not considered how these vary with the  parameters (i.e. $\alpha$ and $\beta$) or the form of the potential itself. Moreover, while our collisions are sub-relativistic they are still faster than their Newtonian escape velocity so it will be worthwhile to examine  interactions between mismatched oscillons depends on their speed.  

As  pointed out in Ref.~\cite{Amin:2011hj}  bounds on the amplitude of an inflationary gravitational wave background suggest that the inflationary potential is sub-quadratic, a necessary condition for oscillon formation. Consequently, it is increasingly plausible that the early universe  passed through an oscillon dominated phase and this investigation is a precursor to a full understanding of the dynamics of any post-inflationary oscillon-dominated era. 

In particular, oscillons persist for timescales comparable to the post-inflationary Hubble time.  Oscillon-oscillon gravitational interactions are suppressed in rigid spacetime backgrounds and  oscillons are fixed in space. However, local gravitational effects could drive oscillon clustering and induce interactions and these are largely unexplored.  Interestingly, there are simulations of oscillon formation in full General Relativity \cite{Kou:2019bbc} but not (so far as we are aware) in the Newtonian limit. However, this would be more computationally tractable for longer simulations and akin to numerical treatments of gravitationally driven structure formation in the present epoch or  the early universe  \cite{Amin:2019ums,Musoke:2019ima,Eggemeier:2020zeg,Eggemeier:2021smj}. 
  
If oscillons do cluster gravitationally the drag discussed in Section 4 would tend to damp lateral motion between pairs as they accelerate towards each other, increasingly the likelihood of collisions. However, gravitationally bound pairs or clusters of oscillons are presumably a possibility and these are are seen in the Gross-Pitaeveskii simulations (that is,  Schr\"{o}dinger-Poisson with self-interacting  matter) performed by Amin and Mocz \cite{Amin:2019ums}. More generally, it will be interesting to understand the extent to  which Gross-Pitaeveskii simulations match solutions to the  Klein-Gordon equation with local gravity. 

It will be important to understand how the existence and properties of any oscillon dominated phase varies between candidate inflationary models and within the  parameter space of each model. A period of oscillon domination will influence the post-inflationary equation of state which in turn modifies the expected values of the inflationary observables \cite{Dodelson:2003vq,%
 Liddle:2003as,Adshead:2010mc,Munoz:2014eqa}. Likewise both resonance itself and oscillon formation more specifically can source high frequency gravitational waves \cite{Zhou:2013tsa} and the form of any such spectrum depend on the details of this era. Lastly, the physics of reheating and thermalization must be understood in the context of any previous oscillon dominated era. 

\vspace{5mm}
 \mbox{}

\begin{acknowledgments}
We are grateful to Leon Southey-Ray for  conversations during the course of this work and to Mustafa Amin for very useful commentary on a draft of the paper.  We acknowledge support from the Marsden Fund managed through Royal Society Te Ap\=arangi. This project utilized NeSI high-performance computing facilities.
\end{acknowledgments}

\bibliography{refs}

\end{document}